\documentclass[10pt,comsoc]{IEEEtran} %conference
\usepackage{amsfonts}
\usepackage{times}
\usepackage{graphicx}
\usepackage{graphicx}
\graphicspath{ {./figures} }
\usepackage{latexsym}
\usepackage{dsfont}
\usepackage{amssymb}
\usepackage{amsmath}
\usepackage{cite}
\usepackage{verbatim}
\usepackage{subfigure}

\usepackage{multirow}
\usepackage{amsmath}  % For math symbols
\usepackage{cite}
\usepackage{longtable}  % For long tables that span multiple pages
\usepackage{makecell}  % To allow line breaks in table cells
\def\bb0{{\mathbb{0}}}

\def\bb{{\mathbf{b}}}

\def\b0{{\mathbf{0}}}

\def\sf0{{\mathsf{0}}}

\usepackage{epstopdf}
\usepackage{enumerate}
\usepackage{algorithmicx}
\usepackage{algorithm}
\usepackage{amsmath}
\usepackage[noend]{algpseudocode}
\usepackage{float}
\usepackage{hyperref}
\usepackage{color}
\usepackage{makeidx}
\usepackage{bbm}
\usepackage{graphicx}
\usepackage{booktabs}
\usepackage{subfigure}
\usepackage{multirow}

\usepackage{stackengine} 
\stackMath

\newlength\matfield
\newlength\tmplength

\usepackage{relsize}

\begin{document}
\title{A Dataset Similarity Evaluation Framework for Wireless Communications and Sensing}

\author{João Morais$^1$, Sadjad Alikhani$^1$, Akshay Malhotra$^2$, Shahab Hamidi-Rad$^2$, Ahmed Alkhateeb$^1$ \\ 
$^1$\{joao, alikhani, alkhateeb\}@asu.edu , $^2$\{akshay.malhotra, shahab.hamidi-rad\}@interdigital.com }
\maketitle

% Original abstract
%\begin{abstract}
%In wireless communications and sensing, machine learning models are often trained on synthetic datasets that may not fully capture the complexities of real-world environments, resulting in an overestimation of model performance. The significant discrepancies between synthetic and real-world datasets highlight the need for a robust method to quantify these differences and their impact on model generalization. This paper introduces a task-specific, model-agnostic metric for evaluating dataset similarity, providing a means to assess and compare dataset realism and quality. Such a metric is crucial for augmenting real-world data, improving benchmarking, and making informed retraining decisions when adapting to new deployment settings, such as different sites or frequency bands.
%\end{abstract}

% Last min change requested by prof... 
\begin{abstract}
This paper introduces a task-specific, model-agnostic framework for evaluating dataset similarity, providing a means to assess and compare dataset realism and quality. Such a framework is crucial for augmenting real-world data, improving benchmarking, and making informed retraining decisions when adapting to new deployment settings, such as different sites or frequency bands. The proposed framework is employed to design metrics based on UMAP topology-preserving dimensionality reduction, leveraging Wasserstein and Euclidean distances on latent space KNN clusters. The designed metrics show correlations above 0.85 between dataset distances and model performances on a channel state information compression unsupervised machine learning task leveraging autoencoder architectures. The results show that the designed metrics outperform traditional methods.
\end{abstract}

\vspace{-.2cm}

\section{Introduction}

Machine learning (ML) applications in wireless communications have seen significant growth, driven by the need for enhanced spectral efficiency and system optimization \cite{8322184, 9370097, 8395149, 9121328, 8938771}. While considerable attention has been given to developing advanced learning models, there has been less focus on the data required for training and generalization. This imbalance limits the transition of ML models from research to real-world deployment in wireless systems. To address this, several key challenges must be resolved:

\begin{itemize}
    \item How to choose adequate datasets for training models?
    \item How to predict model performance in real deployments?
    \item How to measure and ensure model generalization across different datasets?
\end{itemize}

This work delves into the development and application of \textbf{dataset similarity metrics} in the context of wireless communications. These metrics quantify the degree of similarity between datasets, allowing for the estimation of a model’s generalization performance on unseen data without requiring explicit retraining. Beyond generalization prediction, dataset similarity metrics are also useful for detecting distributional shifts, improving transfer learning by selecting the most relevant datasets, and augmenting real-world data with synthetic datasets that are well-matched to the task. In the domain of wireless communications, these metrics offer valuable insights into how datasets relate to one another, guiding efficient model selection and enhancing performance, particularly in scenarios where labeled data is limited or unavailable.

\iffalse
These dataset similarity metrics can be categorized into four main types: geometric distances (e.g., Euclidean, cosine) \cite{liberti2012euclideandistancegeometryapplications, XIA201539}, which measure spatial relationships between data points; statistical distances (e.g., KL divergence, Wasserstein) \cite{10.1214/aoms/1177729694, Panaretos_2019}, used to compare probability distributions; subspace distances (e.g., Grassmannian, principal angles) \cite{10.1145/1390156.1390204, ye2016schubertvarietiesdistancessubspaces}, which assess relationships between subspaces in high-dimensional data; and manifold-based distances (e.g., geodesic, diffusion) \cite{umap_paper, tsne_paper}, which capture relationships in nonlinear spaces. These metrics are critical for understanding dataset structure and improving model generalization in wireless environments.
\fi

In wireless communications, obtaining real-world data is challenging, making simulated datasets essential for ML development. Simulations are getting increasingly realistic, and some studies already permit capturing real world environments by finding the right materials via channel measurements \cite{jiang2024learnablewirelessdigitaltwins}. However, the simulated datasets of wireless channels are often complex to interpret and manage. This highlights the need for robust data engineering practices, guiding the entire data lifecycle from acquisition to transformation and management. Despite the advances in advanced learning techniques, such as the Large Wireless Models \cite{alikhani2024largewirelessmodellwm}, the lack of tools for dataset management is still responsible for slowing the development of large-scale ML models in wireless systems.

%Real-world datasets, such as those from NYU and PAWR, provide valuable insights but are often insufficient in scale and diversity for comprehensive ML model development. To address this, simulated data plays a crucial role. Stochastic models, like those defined by 3GPP’s 38.901 (e.g., CDL, TDL), simulate channel conditions probabilistically, while deterministic approaches, such as ray tracing (e.g., Wireless InSite, SionnaRT), offer site-specific, high-fidelity simulations. Datasets like DeepMIMO, which blend real and synthetic data, further bridge this gap. Despite these advancements, the wireless community still lacks comprehensive methods to assess and manage datasets for large-scale generative models.
Real-world datasets, such as those from OTA measurements by NYU \cite{akdeniz2014millimeter} and testbeds like PAWR \cite{pawr}, offer valuable insights but are limited in scale and diversity for machine learning model training. To address this, simulated data is widely used, falling into two main categories: stochastic and deterministic. Stochastic models, such as 3GPP's 38.901 specification \cite{3GPP_38901} (e.g., CDL, TDL, UMa, UMi), simulate probabilistic channel conditions and are accessible via tools like Matlab’s 5G Toolbox. Other models include NYUSIM \cite{sun2017novel} and tools by Fraunhofer \cite{fraunhofer2023annualreport}. Deterministic approaches, such as ray tracing tools like Wireless InSite \cite{wireless_insite} and SionnaRT \cite{sionna_rt}, provide high-fidelity, site-specific simulations. Datasets like DeepMIMO \cite{alkhateeb2019deepmimo} combine real and synthetic data to bridge this gap. However, the wireless community still lacks robust methods to assess and manage datasets for large-scale generative models, which are crucial for improving model generalization and real-world deployment.

To overcome these limitations, leveraging existing datasets that are distributionally similar to target environments can reduce the need for creating new data from scratch. This highlights the importance of measuring dataset similarity to ensure effective augmentation and improve model performance.

\begin{figure}[t]
    \centerline{\includegraphics[width=1\columnwidth]{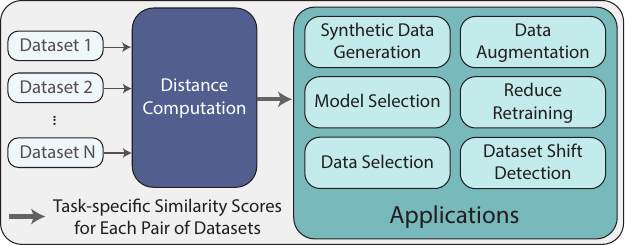}}
    \caption{Applications enabled by dataset distance computation.}
    \label{fig:applications}
\end{figure}

\begin{figure*} [t]
 	\centerline{\includegraphics[width=2\columnwidth]{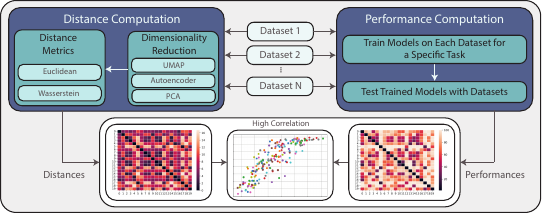}}
 	\caption{Framework system for assessing the suitability of a distance function to task, model, and a set of datasets in terms of how such a function outputs distances that correlate with performances in the specific task.}
 	\label{fig:system_model}
 \end{figure*} 
 
\textbf{Contribution:} This work introduces a framework for assessing dataset similarity, allowing researchers to evaluate datasets before training and determine the need for retraining, among several other application, as shown in Fig. \ref{fig:applications}. The main contributions are:

\begin{itemize}
    \item We develop a task-driven, model-agnostic framework that evaluates similarity between datasets, without the need for training additional models. 
    \item We design two distance metrics using UMAP topology-preserving dimensionality reduction: One computes the Euclidean distances on KNN clusters, the other calculates the Wasserstein distances across data dimensions. 
    \item We demonstrate that our framework achieves a strong correlation between dataset distances and model performances, offering the potential to bypass redundant computations on new datasets and allowing more effective and efficient model training, among several other applications.
\end{itemize}

The proposed framework is open-source, with full documentation and reproducibility resources available. \footnote{Documentation, artifacts, and reproducibility resources can be found at: \textit{https://wi-lab.net/research/dataset\_similarity}} 

\section{Framework for Evaluating Dataset Similarities and Model Performance Correlation} \label{sec:framework}

We propose a framework to correlate dataset distances with model performance metrics, enabling the selection of distance functions that predict how well models trained on one dataset perform on others, depicted in Fig. \ref{fig:system_model}. This framework can detect dataset shifts, guide data augmentation, and rank the usefulness of datasets for specific tasks.

\textbf{Framework Overview:} The framework consists of two steps: distance computation and performance evaluation. For distance computation, a metric \( d \) is calculated between pairs of datasets, resulting in a distance matrix \( \mathbf{D} \). Performance evaluation involves training a model on one dataset and testing it on others, generating a performance matrix \( \mathbf{P} \). Correlating \( \mathbf{D} \) and \( \mathbf{P} \) identifies how well distances predict performance drops between datasets.

\textbf{Performance Computation:} Models are trained on each dataset \( D_i \), and their performance \( P_{ii} \) is evaluated on the same dataset as a baseline. The trained model is then tested on other datasets \( D_j \) to record performance \( P_{ij} \). If \( D_i \) and \( D_j \) are similar based on the task, \( P_{ij} \) should be close to \( P_{ii} \); significant performance drops \( \Delta P_{ij} \) indicate dataset dissimilarity.

\textbf{Distance Computation:} Various distance metrics, including geometric, statistical, subspace, and manifold distances, as well as those based on dimensionality reduction (e.g., UMAP), can be applied within our framework. High-dimensional datasets present challenges due to the curse of dimensionality, but techniques like PCA and UMAP can project data into lower-dimensional spaces, making distances more meaningful and computationally efficient. Our framework is flexible, allowing preprocessing steps like dimensionality reduction or clustering to be adapted based on the task and dataset characteristics.

\textbf{Correlation Between Distance and Performance:} We use the Pearson correlation coefficient to quantify the relationship between the dataset distance matrix \( \mathbf{D} \) and the performance drop matrix \( \mathbf{P} \). This helps us evaluate how well distances predict model performance degradation across datasets. The goal is to find distance metrics that align closely with model behavior, enabling more effective data selection and transfer learning strategies. By applying this framework to unsupervised tasks, we aim to identify effective distance metrics that generalize across domains.

\section{Dataset Similarity: A Novel Approach}

In high-dimensional datasets, computing distances that accurately reflect the relationships between datasets and correlate with model performance is a challenging task, largely due to the presence of noise, irrelevant features, and the complexity of the data. Traditional distance metrics, when applied in their native high-dimensional space, frequently fail to capture the crucial underlying structures. To address this, it is essential to map the datasets into a transformed space that emphasizes the most relevant features for the task, preserving \textbf{local proximity} (datasets close in terms of task-relevant features remain close) and \textbf{global structure} (broader relationships between datasets are maintained). This transformed space allows us to compute distances that are more meaningful, leading to higher correlations with model performance across datasets.

We can achieve this transformation using a \textbf{graph-based approach}, where the relationships between datasets are modeled based on their local neighborhoods and global connections. This method creates a \textbf{manifold-like representation}, reflecting both spatial proximity and structural properties, allowing distances to better represent model performance across datasets, ultimately improving transfer learning, domain adaptation, and model selection.

\subsection{Uniform Manifold Approximation and Projection (UMAP)}

The key tool for this transformation is \textbf{Uniform Manifold Approximation and Projection (UMAP)} \cite{umap_paper}. Let \( D_i = \{ \mathbf{x}_j^{(i)} \}_{j=1}^{M_i} \) be a dataset with \( M_i \) datapoints, where \( \mathbf{x}_j^{(i)} \in \mathbb{R}^N \). UMAP projects this dataset into a lower-dimensional space, represented as \( \tilde{D}_i = \{ \tilde{\mathbf{x}}_j^{(i)} = f_{\text{UMAP}}(\mathbf{x}_j^{(i)}) \}_{j=1}^{M_i} \), where the distances better capture local and global structures compared to the raw feature space. UMAP is a dimensionality reduction technique that can be used for general non-linear dimensionality reduction. The algorithm relies on three key assumptions about the data: i) The data is uniformly distributed on a Riemannian manifold, ii) the Riemannian metric is locally constant (or can be approximated as such), iii) the manifold is locally connected. From these assumptions, it is possible to model the manifold with a fuzzy topological structure. UMAP finds a low-dimensional projection of the data that preserves this fuzzy topological structure as closely as possible. The process is provided below.

\begin{itemize}
    \item \textbf{Constructing a Fuzzy Simplicial Set}: UMAP builds a weighted k-nearest neighbor graph in the high-dimensional space, capturing local neighborhood relationships based on connection probabilities between points.
    \item \textbf{Fuzzy Topological Representation}: These probabilities are used to construct a topological space, reflecting both local and global structures of the dataset.
    \item \textbf{Low-Dimensional Embedding Optimization}: UMAP finds a low-dimensional embedding that minimizes the cross-entropy between fuzzy representations in the high-dimensional and low-dimensional spaces.
\end{itemize}

Preserving both local and global relationships, UMAP provides a better alternative to PCA (which struggles with non-linear structures) and t-SNE \cite{tsne_paper} (which distorts global structures). This makes it ideal for computing distances that reflect how models trained on one dataset perform on another.

\subsection{Dataset Similarity Metrics in UMAP Spaces}

\textbf{Euclidean in UMAP Spaces}: In the UMAP-encoded latent space, various forms of Euclidean distances are utilized to quantify dataset separation, providing computational simplicity and interpretability:

\textit{Pairwise Euclidean Distance}: The distance between all pairs of points across two datasets \( D_1 \) and \( D_2 \), where \( M_1 \) and \( M_2 \) are the number of data points in the datasets, is given by
    \[
    d_{\text{pairwise}} = \frac{1}{M_1 M_2} \sum_{j=1}^{M_1} \sum_{k=1}^{M_2} \left\| \tilde{\mathbf{x}}_j^{(1)} - \tilde{\mathbf{x}}_k^{(2)} \right\|_2.
    \]
    Here, \( \tilde{\mathbf{x}}_j^{(1)} \) and \( \tilde{\mathbf{x}}_k^{(2)} \) represent the encoded points from datasets \( D_1 \) and \( D_2 \), respectively.

\textit{Cluster-Based Euclidean Distance}: After clustering each dataset into \( K_1 \) and \( K_2 \) clusters, the centroid-based Euclidean distance is computed as
    \[
    d_{\text{cluster}} = \frac{1}{K_1 K_2} \sum_{l=1}^{K_1} \sum_{m=1}^{K_2} \left\| \mathbf{c}_l^{(1)} - \mathbf{c}_m^{(2)} \right\|_2,
    \]
    where \( \mathbf{c}_l^{(1)} \) and \( \mathbf{c}_m^{(2)} \) are the centroids of clusters \( l \) and \( m \) in datasets \( D_1 \) and \( D_2 \), respectively.

\textit{Average Euclidean Distance}: This distance simplifies to the Euclidean distance between the mean vectors \( \bar{\mathbf{x}}^{(1)} \) and \( \bar{\mathbf{x}}^{(2)} \) of the two datasets
    \[
    d_{\text{average}} = \left\| \bar{\mathbf{x}}^{(1)} - \bar{\mathbf{x}}^{(2)} \right\|_2,
    \]
    where \( \bar{\mathbf{x}}^{(1)} = \frac{1}{M_1} \sum_{j=1}^{M_1} \tilde{\mathbf{x}}_j^{(1)} \) and similarly for \( \bar{\mathbf{x}}^{(2)} \).

The advantage of Euclidean distances in UMAP spaces lies in their computational efficiency and ability to clearly reflect local dataset structure after dimensionality reduction, which is crucial for fast and intuitive dataset similarity evaluation.

\textbf{Wasserstein in UMAP Spaces}: The Wasserstein distance \cite{villani2008optimal, peyre2019computational, earthmover}, also known as the Earth Mover's distance, is a more powerful metric for measuring differences between two distributions. It calculates the minimal effort needed to transport the mass of one distribution to match the other across all dimensions of the latent space. For each dimension \( n \), the one-dimensional Wasserstein distance between the cumulative distribution functions (CDFs) \( F_n^{(1)} \) and \( F_n^{(2)} \) of datasets \( D_1 \) and \( D_2 \) is
\[
W_n = \int_{0}^{1} \left| F_n^{(1)^{-1}}(t) - F_n^{(2)^{-1}}(t) \right| dt,
\]
where \( F_n^{(i)^{-1}}(t) \) is the inverse CDF (quantile function) for the \( n \)-th dimension of dataset \( D_i \).

The overall Wasserstein distance across all dimensions \( d \) of the latent space is then the average of these one-dimensional Wasserstein distances as follows
\[
W = \frac{1}{d} \sum_{n=1}^{d} W_n.
\]

Wasserstein distance is especially advantageous for capturing the global structure of datasets, providing superior sensitivity to both local and global distributional differences. This makes it highly effective in UMAP spaces, where preserving both local and global features is critical for accurate dataset similarity evaluation.

\textbf{UMAP Considerations}: 
UMAP requires careful tuning of parameters, such as the number of neighbors and minimum distance in the latent space. A balance between local and global structures must be achieved for effective embedding. For wireless datasets, correlation distance has proven effective for UMAP, but Euclidean distance can be used when scalability is prioritized over performance. Leveraging UMAP for distance computation offers improved accuracy in assessing dataset similarities, providing a robust foundation for dataset distancing tasks.

\begin{figure*}[t]
    \centerline{\includegraphics[width=2\columnwidth]{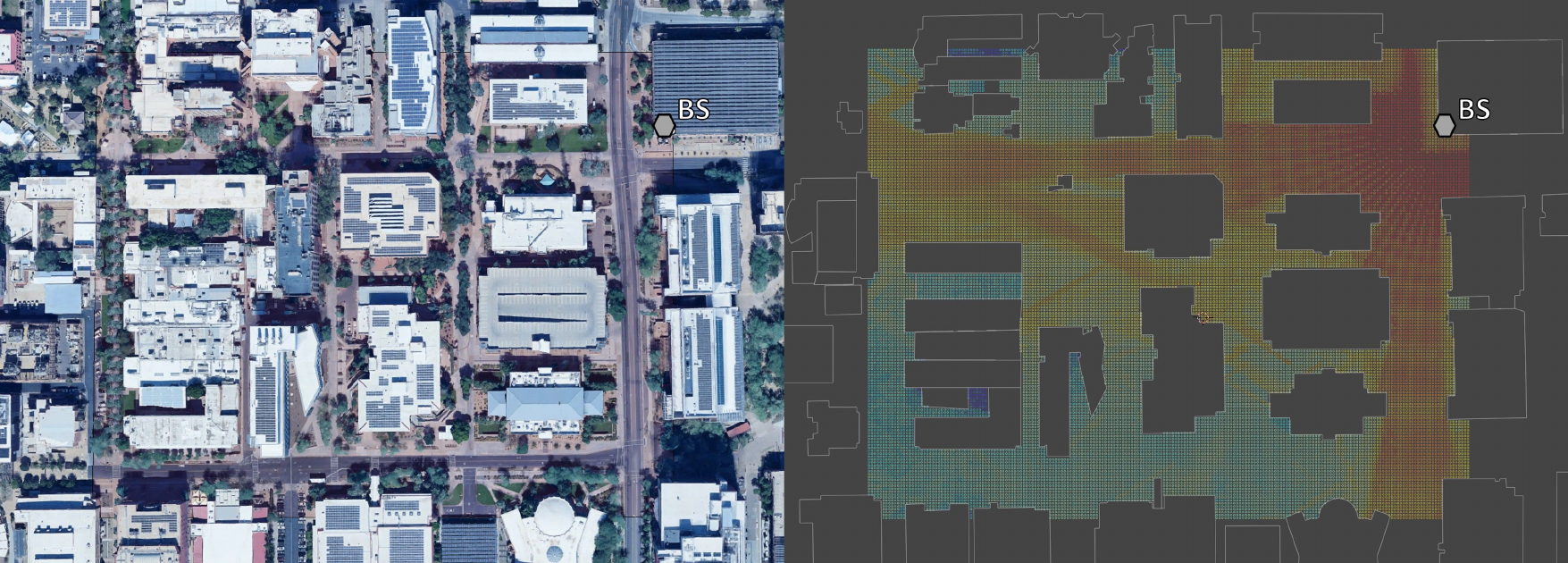}} 
    \caption{Real (left) and rendered (right) top views of the ASU campus from the DeepMIMO dataset. The rendered view shows received power distribution using a standard DFT codebook at the base station, highlighting key effects like roof diffraction.}
    \label{fig:dataset}
\end{figure*}

\section{Dataset Similarity For CSI Compression Task} \label{sec:unsup}

This section explores how various distance metrics correlate with model performance in the unsupervised CSI compression task, which is essential in wireless communications. CSI compression reduces high-dimensional channel matrices into low-dimensional representations to facilitate efficient feedback between user equipments and base stations. Since the same data serves as both input and output, this task is ideal for assessing correlations between dataset distances and model performance. We discuss the CSI compression task, the autoencoder model, and the dataset before presenting results of different distance metrics in both raw and latent spaces.

\subsection{CSI Compression Task}

CSI compression involves transforming a high-dimensional wireless channel matrix, $\mathbf{H} \in \mathbb{C}^{N_{BS} \times N_{sub}}$, into a compact, low-dimensional representation. Here, we consider $N_{BS}$ basestation antennas and $N_{sub}$ subcarriers in the channel matrix. After applying a Fourier transform and truncating the last 16 delay taps, the matrix is reduced to $32 \times 16$. The goal is to compress this matrix to $N_{enc} = 32$ dimensions, achieving a $64$-fold reduction. The compression performance is measured using the normalized mean squared error (NMSE) between the original matrix $\mathbf{H}$ and the reconstruction $\hat{\mathbf{H}}$ as follows
\[
    \text{NMSE}_{\text{dB}}(\mathbf{H}, \hat{\mathbf{H}}) = 10 \log_{10} \frac{\|\mathbf{H} - \hat{\mathbf{H}}\|^2_F}{\|\mathbf{H}\|^2_F}.
\]

\subsection{Autoencoder Model}

\begin{figure*} [t]
	\centerline{\includegraphics[width=2\columnwidth]{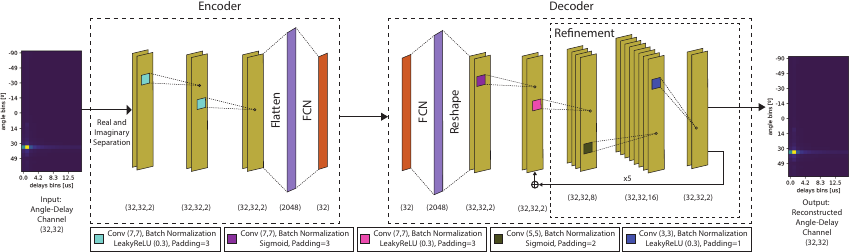}}
	\caption{Architecture of the autoencoder model used for the unsupervised CSI compression task, inspired by CSINet+ \cite{guo2019convolutionalneuralnetworkbased}.}
	\label{fig:AE}
\end{figure*}

The autoencoder (AE) used for CSI compression processes the input $32 \times 16$ matrix, split into real and imaginary parts, through convolutional layers to reduce it to a $32$-dimensional latent space. The decoder reconstructs the high-dimensional matrix from the latent space. The model is trained using MSE loss and is configured in two ways:
\begin{itemize}
    \item AEs are trained separately on each area dataset. Each model achieves an NMSE below $-20 \text{dB}$ when tested on the trained area and higher NMSE on untrained areas. These 20 models are used to access performance drops that should be correlated with our distance metrics.
    \item A single AE with five refinement nets (instead of three) is trained on data from all areas, resulting in an average NMSE of $-20 \text{dB}$ across areas. This model latent space is used for computing dataset distances later on when analyzing the impact of dimensionality reduction methods.
\end{itemize}

\subsection{Dataset Description}

The dataset used in this work is a raytraced ASU campus dataset generated with the DeepMIMO framework, containing around $90K$ users across an area of $410 \times 320$ meters with 1-meter resolution. The raytracing simulation captures various propagation effects, including LoS, reflections, diffuse scattering, and diffraction, providing a highly realistic environment. Figure~\ref{fig:dataset} compares the real geographic view of the ASU campus with the digitally rendered version, illustrating the dataset's suitability for evaluating distance metrics in wireless channel compression tasks.

\iffalse
\begin{figure*}[t]
	\centerline{\includegraphics[width=2\columnwidth]{distances.pdf}}
	\caption{Classes of distances that can be applied between datasets. }
	\label{fig:distances}
\end{figure*}
\fi 

\begin{table}[t]
\caption{Correlation between dataset distances and model performance in the input space.}
\label{tab:raw_unsup}
\centering
\begin{tabular}{|l|l|c|c|}
\hline
\textbf{Category} & \textbf{Distance Name}  & \textbf{Correlation} & \begin{tabular}[c]{@{}c@{}}\textbf{Compute}\\ \textbf{Time (s)}\end{tabular} \\ \hline
\multirow{4}{*}{Geometric} & Pairwise Euclidean      & 0.36  & 11  \\  
                           & Clustered Euclidean     & 0.37  & 166 \\  
                           & Average Euclidean      & 0.34  & 3   \\  
                           & Cosine                  & -0.07 & 5255 \\ \hline
\multirow{8}{*}{Statistical} & Jensen-Shannon        & 0.14  & 80  \\  
                           & Hellinger               & 0.15  & 81  \\  
                           & Wasserstein             & 0.52  & 562 \\  
                           & Kolmogorov-Smirnov      & 0.47  & 381 \\  
                           & Total Variation         & 0.15  & 78  \\  
                           & MMD (linear)            & -0.08 & 79  \\  
                           & MMD (RBF)               & -0.06 & 135 \\  
                           & Energy                  & 0.55  & 735 \\ \hline
\multirow{3}{*}{Subspace}  & Grassmann               & -0.11 & 15223 \\  
                           & Chordal                 & -0.10 & 14810 \\  
                           & Asimov                  & -0.03 & 15594 \\ \hline
Other                      & PAD                     & 0.64  & 952   \\ \hline
\end{tabular}
\end{table}

\subsection{Results: Distances in Input Space}

Table \ref{tab:raw_unsup} presents the correlations between different distance metrics and model performance for the CSI compression task in the raw input space. We evaluate three categories of distance metrics: geometric, statistical, and subspace-based. Key findings include:
\begin{itemize}
    \item \textbf{Statistical distances outperform geometric ones.} Wasserstein ($0.52$ correlation) and Energy ($0.55$ correlation) distances perform best, because they capture distributional differences, unlike point-to-point metrics that struggle with high-dimensional data.
    \item \textbf{Geometric distances show lower correlations.} Averaged and pairwise Euclidean distances achieve roughly $0.35$ correlation, due to their limitations for high-dimensional wireless data.
    \item \textbf{Subspace distances performed poorly} Grassmann, Chordal, and Asimov metrics perform the worst and even yield negative correlations. This can be associated with the similarity between projected subspaces, resulting in practically constant principal angles and, thus, distances.
    \item \textbf{Computation time matters.} While Wasserstein and Energy perform well, they are computationally expensive (562s and 735s, respectively). Simpler methods like centroid Euclidean (3s) are faster but offer lower correlations.
\end{itemize}

These results highlight that while statistical distances are more effective, their computational cost motivates the need for dimensionality reduction to make distance computations more practical.

\subsection{UMAP for Improved Distance Computation}

\begin{figure*} [t]
    \centerline{\includegraphics[width=2\columnwidth]{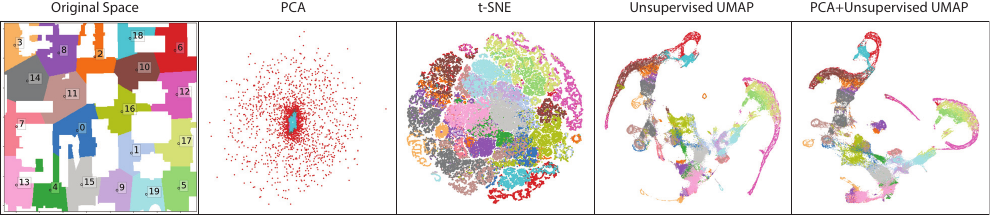}}
    \caption{Visualization of latent spaces generated by different dimensionality reduction techniques. From left to right: the original space with proximity-based clustering, followed by PCA, t-SNE, UMAP, and a combination of PCA and UMAP.}
    \label{fig:unsup_comparison}
\end{figure*}

Dimensionality reduction helps mitigate the high computational costs of raw space distance computation by removing noise and redundancies. Fig.~\ref{fig:unsup_comparison} compares several techniques, including PCA, t-SNE, and UMAP.

UMAP was found to be the most effective among dimensionality reduction techniques, striking a balance between preserving local and global structures. PCA, being linear, fails to differentiate between similar datasets, while t-SNE distorts global relationships by focusing on local clusters. UMAP retains essential geometric features, resulting in a latent space where distances better reflect dataset similarities.

\subsection{Results: Distances in Latent Space}

Table ~\ref{tab:dim_red} summarizes the performance of various distance metrics in different latent spaces, using PCA, t-SNE, UMAP, and autoencoder (AE).

\begin{table}[t]
\setlength{\tabcolsep}{0.4em} % for the horizontal padding
\centering
\caption{Correlation of distances with model performance in different latent spaces.}
\begin{tabular}{|c|l|cccc|}
\hline
\multirow{2}{*}{\textbf{Category}}          & \multirow{2}{*}{\textbf{Distance}} & \multicolumn{4}{c|}{\textbf{Dimensionality Reduction}} \\ %\cline{3-7} 
                                            &                                        & \textbf{PCA 32} & \textbf{TSNE 2} & \textbf{UMAP 2} &  \textbf{AE 32} \\ \hline
\multirow{4}{*}{\textbf{Geometric}}         & Pairwise Euclidean                     & 0.37            & 0.58            & \textbf{0.83}   & \textbf{0.87}    \\ %\cline{2-7} 
                                            & Clustered Euclidean                    & 0.41            & 0.59            & \textbf{0.84}   & \textbf{0.91}    \\ %\cline{2-7} 
                                            & Centroid Euclidean                     & 0.35            & 0.59            & \textbf{0.86}   & \textbf{0.93}    \\ %\cline{2-7} 
                                            & Cosine                                 & 0.30            & 0.41            & 0.42            & \textbf{0.94}    \\ \hline
\multirow{9}{*}{\textbf{Statistical}}       & KL Divergence                          & 0.32            & 0.62            & 0.52            & \textbf{0.85}    \\ %\cline{2-7} 
                                            & Jensen-Shannon                         & -0.08           & 0.15            & 0.12            & 0.07             \\ %\cline{2-7} 
                                            & Hellinger                              & -0.08           & 0.17            & 0.13            & 0.07             \\ %\cline{2-7} 
                                            & Wasserstein                            & 0.47            & 0.68            & \textbf{0.85}   & \textbf{0.92}    \\ %\cline{2-7} 
                                            & Kolmogorov-Smirnov                     & 0.57            & 0.32            & 0.46            & 0.22             \\ %\cline{2-7} 
                                            & Total Variation                        & -0.06           & 0.13            & 0.10            & 0.04             \\ %\cline{2-7} 
                                            & MMD (Linear)                           & -0.17           & 0.10            & 0.04            & 0.04             \\ %\cline{2-7} 
                                            & MMD (RBF)                              & -0.13           & 0.06            & 0.04            & -0.02            \\ %\cline{2-7} 
                                            & Energy                                 & 0.56            & 0.42            & 0.60            & 0.25             \\ \hline
\multirow{3}{*}{\textbf{Subspace}}          & Grassmann                              & -0.14           & 0.02            & -0.22           & -0.05            \\ %\cline{2-7} 
                                            & Chordal                                & -0.14           & 0.02            & -0.22           & -0.05            \\ %\cline{2-7} 
                                            & Asimov                                 & -0.12           & 0.03            & -0.23           & -0.06            \\ \hline
\textbf{Other}                              & PAD                                    & \textbf{0.75}   & 0.68            & 0.71            & 0.66             \\ \hline
\end{tabular} \label{tab:dim_red}
\end{table}

\textbf{AE as an upper bound:} The AE-based latent space achieves the highest correlation ($0.94$) between distances and model performance. This is unsurprising, as the AE architecture used for dimensionality reduction is similar to the model used for performance evaluation. However, the downside of this approach is its impracticality. Training an AE for all datasets is computationally expensive and time-consuming, making it an infeasible solution for real-world applications.

\textbf{UMAP as a practical alternative:} UMAP provides a close approximation to the AE’s performance, with correlations of around 0.85 for both Euclidean and Wasserstein distances. UMAP's ability to balance local and global structures makes it a highly effective dimensionality reduction technique. Importantly, UMAP drastically reduces the computational complexity of distance computations, making it a practical choice for large-scale applications.

\textbf{Euclidean and Wasserstein metrics perform best:} In both the UMAP and AE spaces, Euclidean-based distances and Wasserstein distance achieve the highest correlations. This suggests that these metrics are well-suited to the latent spaces generated by UMAP and AEs, capturing meaningful dataset similarities that correlate with model performance.

\textbf{Dimensionality reduction reduces computational costs:} By projecting the data into lower-dimensional latent spaces, we reduce the computational burden of distance computation. For example, computing Wasserstein distance in the raw space takes 562s, but this time is significantly reduced when using UMAP. This makes UMAP a practical choice for improving both performance and computational efficiency.

The choice of dimensionality reduction method plays a crucial role in improving both the accuracy and efficiency of distance computations. While AE-based latent spaces provide the highest correlation with model performance, UMAP emerges as a more practical alternative, offering strong correlations with much lower computational overhead. 
UMAP, combined with simple Euclidean or Wasserstein distances, achieves correlations of around 0.85 with model performance, providing a close approximation to the upper bound set by the AE. This makes UMAP an attractive choice for real-world applications, where computational efficiency and scalability are critical considerations.

\section{Conclusion} \label{sec:conclusion}

In this work, we introduced a novel framework for dataset similarity evaluation in wireless communications, establishing one of the first links between dataset distances and model performance. By utilizing latent spaces derived from UMAP non-linear dimensionality reduction, we captured essential data structures for precise distance measurements in unsupervised tasks. Our results showed that the proposed metrics outperformed traditional methods, particularly in CSI compression. This framework enables smarter data selection, reduces model retraining, and supports more efficient machine learning deployment in wireless systems. Future work will extend these insights to broader wireless tasks and real-world datasets.

\bibliographystyle{IEEEtran}

\end{document}